# Nonvolatile electrically reconfigurable integrated photonic switch

Jiajiu Zheng[1*], Zhuoran Fang[1], Changming Wu[1], Shifeng Zhu[2], Peipeng Xu[3], Jonathan K. Doylend[4], Sanchit Deshmukh[5], Eric Pop[5], Scott Dunham[1,2], Mo Li[1,2] and Arka Majumdar[1,2*]

[1]Department of Electrical and Computer Engineering, University of Washington, Seattle, WA 98195, USA

[2]Department of Physics, University of Washington, Seattle, WA 98195, USA

[3]Laboratory of Infrared Materials and Devices, Advanced Technology Research Institute, Ningbo University, Ningbo 315211, China

[4]Silicon Photonic Products Division, Intel Corporation, Santa Clara, CA 95054, USA

[5]Department of Electrical Engineering, Stanford University, Stanford, CA 94305, USA

*e-mail: jjzno1@uw.edu; arka@uw.edu

**Reconfigurability of photonic integrated circuits (PICs) has become increasingly important due to the growing demands for electronic-photonic systems on a chip[1] driven by emerging applications, including neuromorphic computing[2], quantum information[3,4], and microwave photonics[5,6]. Success in these fields usually requires highly scalable photonic switching units as essential building blocks[7-9]. Current photonic switches, however, mainly rely on materials with weak, volatile thermo-optic[10] or electro-optic[11,12] modulation effects, resulting in a large footprint and high energy consumption. As a promising alternative, chalcogenide phase-change materials (PCMs) exhibit strong modulation in a static, self-holding fashion[13-17]. Here, we demonstrate nonvolatile electrically reconfigurable photonic switches using PCM-clad silicon waveguides and microring resonators that are intrinsically compact and energy-efficient. With phase transitions actuated by in-situ silicon PIN heaters, near-zero additional loss and reversible switching with high endurance are obtained in a complementary metal-oxide-semiconductor (CMOS)-compatible process. Our work**

**can potentially enable very large-scale general-purpose programmable integrated photonic processors.**

Advances in PICs, in particular, silicon photonics, have enabled chip-scale integration of photonic systems with electronic circuits[1]. As Moore's Law slows down and the electronic von Neumann bottleneck (*i.e.* the information traffic jam between the memory and processor) gets worse, these photonic systems are becoming more and more attractive by offering broad bandwidth with energy-efficient information transport, processing, and storage[15]. They are especially desirable when it comes to emerging applications such as neuromorphic computing[2], quantum information[3,4], and microwave photonics[5,6], most of which require general-purpose programmable PICs[7-9]. Inspired by the field-programmable gate arrays (FPGAs) in electronics, such generic PICs or optical FPGAs usually consist of a mesh of photonic switches that can be reconfigured on demand to the bar, cross, or coupler states[7-9] to provide different functionalities, including universal linear transformation[8,9]. Current switches in these photonic systems, however, primarily rely on thermo-optic[10] or electro-optic[11,12] modulation effects that are weak and volatile. The resulting large footprint and constant high energy consumption thus restrict the scalability of such optical FPGAs.

Chalcogenide PCMs such as $Ge_2Sb_2Te_5$ (GST) have recently raised considerable interest because of their exceptional nonvolatile reconfigurability[13,14]. Upon structural phase transitions between the covalent-bonded amorphous state and the resonant-bonded crystalline state, PCMs exhibit substantial contrast in electrical resistivity[18] and optical constants (usually $\Delta n > 1$) over a broad spectral range[13,14,19]. Once switched, the resulting state can be retained for more than ten years under ambient conditions in no need of any external power supply[18,20]. Additionally, PCMs can be reversibly switched by low-energy[14,21-23] optical or electrical pulses on a sub-nanosecond timescale[24-26] with potentially long endurance up to $10^{15}$ cycles[27]. Besides, PCMs are highly scalable[28] and can be simply deposited via sputtering onto any substrate without the "lattice mismatch" issue. Consequently, PCMs have already been used in compact, low-energy, and broadband programmable PICs for switches[14,16,29-34], memories[15,35], and computing[17,36-38]. However, flexible control of the states of PCMs in a chip-scale PIC system, which will essentially enable scalable photonic switching units as critical building blocks of optical FPGAs, is yet to be resolved.

Previously, reversible phase transitions of PCMs on PICs have been actuated by free-space optical heating, on-chip optical heating, or electrical threshold switching. Free-space optical heating[14,29,30] by far-field focused laser pulses is not viable for large-scale integration due to the slow, diffraction-limited, inaccurate alignment process[14]. On-chip optical heating[15,17,22,23,31,33,35-38], assisted by evanescent coupling of near-field optical pulses from waveguides to PCMs, can support fully integrated all-optical operations but has difficulty in switching large-area PCMs and complex light routing[37], leading to limited switching contrast and system complexity. Both optical heating approaches also suffer from the weak or non-existent photothermal effect of the relatively transparent amorphous state in the re-crystallization process[15,33,39,40]. Whereas classical electrical threshold switching[23] allows large-scale integration, the limited phase transition volume due to the crystallization filamentation[41] and nonuniform heating is not suitable for photonic applications. Recently, electrical switching with external heaters[32-34] has shown promising results in PCM-integrated photonics. However, in these demonstrations, a large insertion loss is incurred due to the use of indium tin oxide (ITO) heaters[32,33] or uniformly doped silicon heaters[34] and the number of switching cycles is limited to ~5-50. Here, we show that, by integrating GST on silicon PIN heaters, it is feasible to reversibly trigger large-area phase transitions over more than 1,000 times (500 cycles) with near-zero additional loss (~0.02 dB/μm). Utilizing GST-on-silicon waveguides and microring resonators, we demonstrate CMOS-compatible, compact, and energy-efficient nonvolatile electrically reconfigurable photonic switching units that exhibit strong attenuation and optical phase modulation effects and are suitable for creating large-scale optical FPGAs.

As illustrated in Fig. 1a, our photonic switching units were fabricated (see Methods) on a silicon-on-insulator (SOI) wafer with a 220-nm-thick silicon top layer. The partially etched silicon waveguide consists of a 500-nm-wide rib on a 100-nm-thick slab layer, supporting the propagation of single-mode transverse electric (TE) light. To form the PIN junction, the slab is heavily doped by boron and phosphorus ion implantation, 200 nm away from the left and right edge of the rib in the active region, respectively. The separation distance is close enough to the rib to reduce the resistance, but still far enough from the TE mode distribution to ensure a negligible optical loss (see Supplementary Section 2 for insertion loss analysis). The doping concentration is chosen as $10^{20}$ cm$^{-3}$ to achieve high conductivity as well as Ohmic contact with the Ti/Pd (5 nm/180 nm) signal (S) and ground (G) electrodes for low-voltage operations. Pd is adopted because of its high temperature resistance. After metallization, a 10-nm or 20-nm thin-film of GST patch was sputtered onto the rib, inducing strong mode modification upon phase transitions through evanescent coupling. The GST and part of the electrodes near the heating region are encapsulated by 30-nm-thick $Al_2O_3$ through atomic layer deposition (ALD) (Fig. 1a-c) to avoid oxidation and prevent the melted GST from reflowing and deforming during the amorphization process[33], thus allowing high cyclability of the device. To obtain uniform heating and a large alignment tolerance in electron-beam lithography (EBL), the width and the length (in the direction of light propagation) of the GST patch are set to be smaller than those of the rib in the active region by 50 nm on each side (Figs. 1a and 1d).

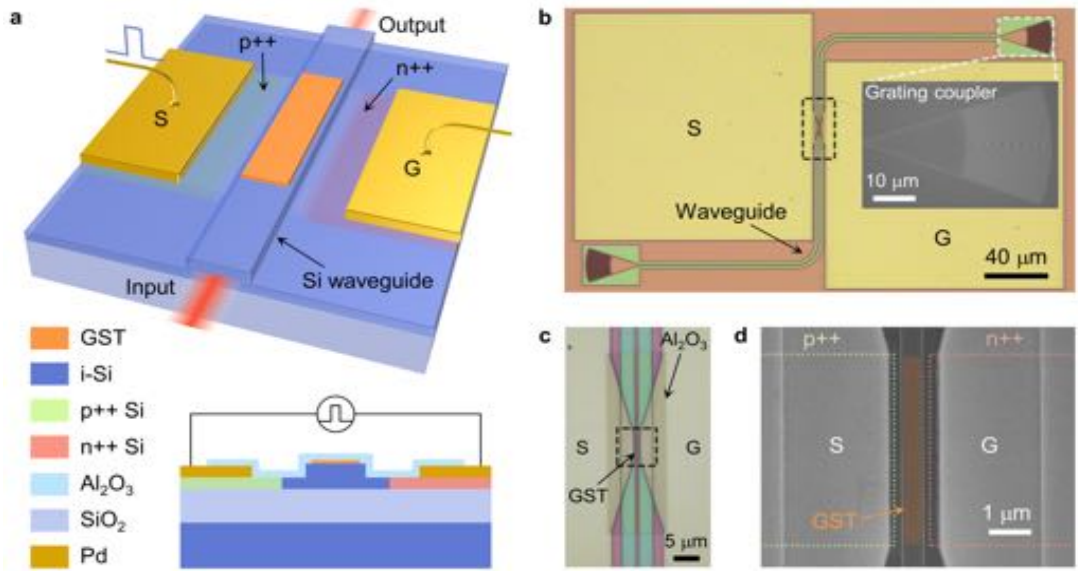

**Figure 1. Nonvolatile electrically reconfigurable photonic switching units. a,** Schematic of the device. For clarity, the top thin-film $Al_2O_3$ encapsulating layer is not displayed. Inset, cross-section of the device. **b,** Top-view optical microscope image of the switching unit on a waveguide with 10-nm-thick GST and a 5-μm-long active region. Inset, SEM image of the grating coupler. **c,** Optical microscope image of the black dashed area in **b**. **d,** SEM image of the active region boxed in **c**.

False color is used to highlight the GST (orange). S (G), signal (ground) electrode. p++ (n++), heavily doped p (n)-type silicon region. i, intrinsic silicon region.

In contrast to optical heating and electrical threshold switching, phase transitions of our photonic switching units rely on the transfer of the electrical pulse-generated Joule heat from the in-situ silicon waveguide PIN junctions. Hence, the switching region can be locally selected and be arbitrarily extended by increasing the size of the heaters, enabling large-scale integration and large-area functional devices such as directional couplers[16]. A 5-µm-long switching unit (*i.e.* the length of the GST patch is 4.9 µm) was operated here to demonstrate the transient response of the optical transmission and the simulated temperature distributions for the crystallization (Set) and amorphization (Reset) processes (see Methods). For the Set process (Fig. 2a), we applied a single pulse of 3.5 V (~10 mW) for 50 µs with a long falling edge of 30 µs (12-ns rising edge for all the pulses in this paper) to ensure that the GST was heated to just above its glass transition temperature ($T_g$) but below the melting point ($T_m$) for a long enough time. The nucleation of small crystallites and subsequent growth can, therefore, proceed, leading to an elevated refractive index ($n$) and extinction coefficient ($\kappa$) of the GST (Supplementary Section 4) and a reduced optical transmission of the device after the pulse. For the Reset process (Fig. 2b), a single pulse of 6.6 V (~110 mW) for 100 ns with a short falling edge of 12 ns (same for all the other Reset pulses) was utilized to melt the GST and then rapidly quench it below $T_g$, forming the disordered glass state with low optical constants and increased optical transmission of the device after the pulse. During the phase transitions, an ultrafast free-carrier absorption effect[42] due to the carrier injection into the silicon was observed from the steep change in the transmission at the sharp edges of the pulses. The slowly changed optical transmission at other times is, however, dominated by the thermo-optic effect of GST[43] in response to the heating and cooling processes. The overall switching period is then determined by the pulse width as well as the dead time due to the thermal relaxation (See Supplementary Section 3 for heating dynamics analysis).

Strong mode modification is expected after the phase transitions according to the simulated mode profiles with amorphous and crystalline GST (see Fig. 2c and Methods), implying substantial absorptive and refractive modulation effects of our switching units.

As the mode distribution is mostly confined in the transparent intrinsic region, near-zero additional loss (~0.02 dB/μm) is introduced by the PIN junction (Supplementary Section 2), showing promising scalability for optical FPGAs. To verify the phase transition processes, we further calculated (see Methods and Supplementary Section 5) the real-time optical transmission due to the effective index change based on the simulated carrier density and temperature distributions (Fig. 2d), exhibiting good qualitative agreement with the experiment.

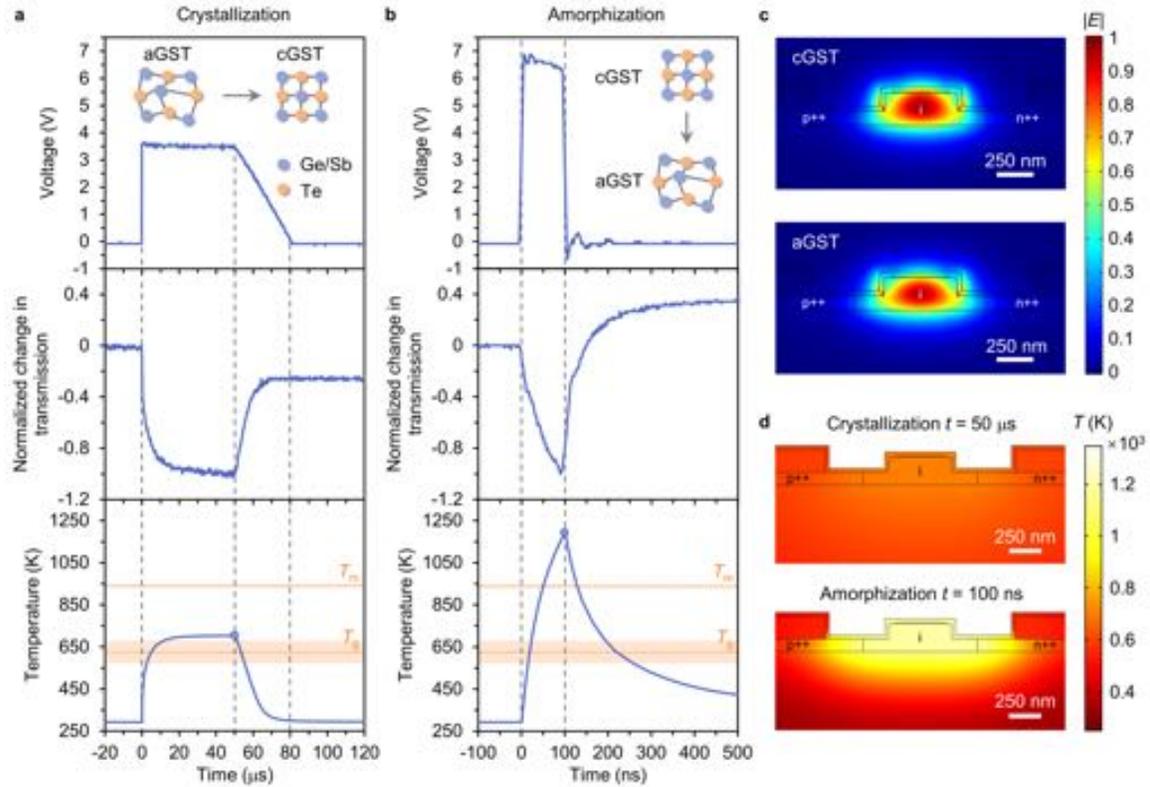

**Figure 2. Operating principle of the photonic switching units. a,b,** Real-time voltage of the applied electrical pulse (upper panel), corresponding change in optical transmission (with the minimum value normalized to –1) of the switching unit at 1,550 nm (middle panel), and simulated temperature response in the center of the GST cross-section to the equivalent pulse with the same power and width (lower panel) for crystallization (Set, **a**) and amorphization (Reset, **b**). Insets of the upper panels, schematic of the GST structure changes due to the phase transitions. The orange shaded areas in the lower panel represent the assumed temperature intervals where the properties of GST are weighted sums of those in the amorphous and crystalline states during phase transitions. $T_g$, glass transition temperature of GST. $T_m$, melting point of GST. **c,** Simulated electric field ($|E|$)

profiles for the fundamental quasi-TE mode of the switching unit at 1,550 nm in the crystalline (top, $n_{eff}$ = 2.68 − 0.05i) and amorphous (bottom, $n_{eff}$ = 2.60 − 1.06 × 10$^{-3}$i) states. **d,** Simulated temperature (*T*) distributions of the switching unit cross-section for crystallization (top) and amorphization (bottom) at the time (*t*) marked by the dots in the lower panels of **a** and **b**, respectively. aGST (cGST), amorphous (crystalline) GST. p++ (n++), heavily doped p (n)-type silicon region. i, intrinsic silicon region. Here, the length of the switching unit is 5 μm and the thickness of the GST is 10 nm.

The Set process can also be actuated using current sweeps. Figure 3a presents the current-voltage (I-V) characteristic of a photonic switching unit on a waveguide during and after the Set by a current sweep, showing a typical rectification behavior of PIN junctions. In contrast to electrical threshold switching[21], no abrupt change of the resistivity is observed in the I-V curves, confirming a different switching mechanism (*i.e.* electrical switching by external heaters) as the GST is not part of the electric circuit due to the relatively large resistance. The corresponding optical output under a current sweep (Fig. 3b), however, does exhibit an obvious change due to the phase transition. Additionally, we find a transient gradual reduction of the output during the current sweep that can be again attributed to the carrier-injection induced loss from silicon and the thermo-optic effect of GST. Manually applying a current sweep (0-16 mA) and a Reset pulse (a single pulse of 4.3 V for 100 ns) shows that the device can be reversibly switched with a high extinction ratio of ~5 dB (1.25 dB/μm for 20-nm-thick GST) over a broad spectral range (Fig. 3c). It is notable that due to the nonvolatility of GST, the output spectra are self-held, precluding any need for external control after the phase transitions. This leads to highly energy-efficient operations compared with thermo-optic or electro-optic switches.

To further inspect the state retention and the cyclability of the switching units on waveguides, we applied numerous Set and Reset pulses and recorded the temporal trace (Fig. 3d) and static states (Fig. 3e) of the transmission change. The results show that our devices allow stable binary operations with reversible phase transitions over more than 1,000 times. No obvious performance degradation was found after repeating the same experiment (1,000 transitions) for several times, implying that a significantly longer endurance can be expected. The transmission fluctuation in Fig. 3d is primarily caused by the gradual misalignment of the setup (see Methods and Supplementary Section 1) during the measurement while a relatively large uncertainty of the transmission in the crystalline state (Fig. 3e) can be explained due to partial crystallization of the GST.

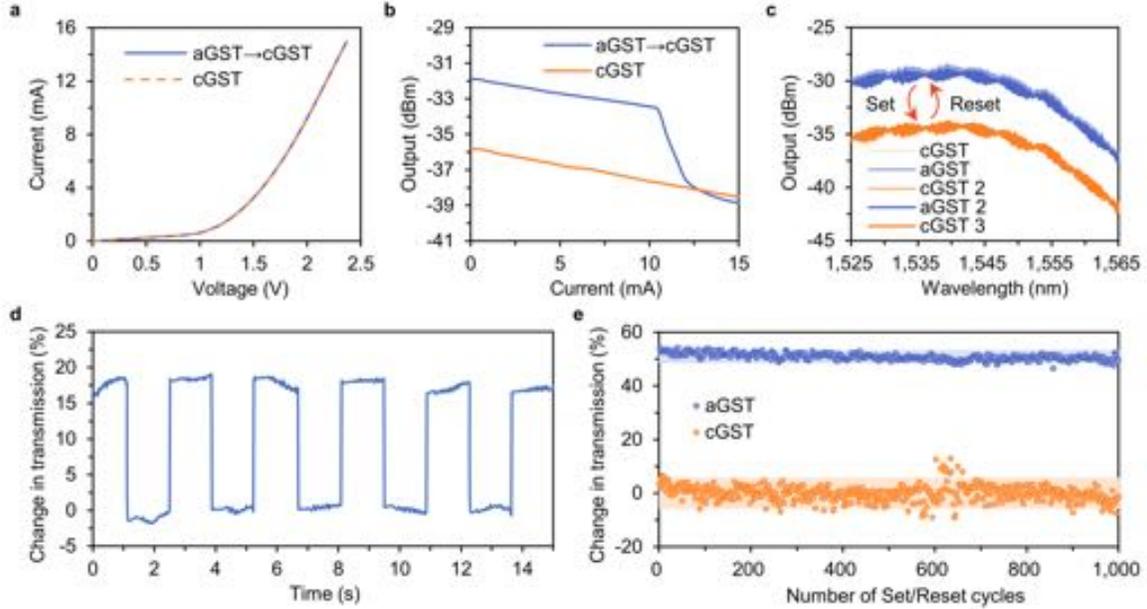

**Figure 3. Performance of the photonic switching units on waveguides. a,** Current-voltage (I-V) curves of the photonic switch with 20-nm-thick GST and a 4-μm-long active region obtained via current sweeps (0-16 mA) during and after the Set process, exhibiting a typical rectification behavior of PIN diodes without the electrical threshold switching effect. **b,** Corresponding optical output under current sweeps at 1,550 nm, showing an abrupt change during the Set process. **c,** Optical output spectra of the same device after two reversible Set (0-16 mA sweep) and Reset (a single pulse of 4.3 V for 100 ns) processes. **d,** Temporal trace of the transmission change in a photonic switch with 10-nm-thick GST and a 4-μm-long active region at 1,550 nm during the consecutive Set (a single pulse of 3.4 V for 1 ms with a falling edge of 0.6 ms) and Reset (a single pulse of 7.1 V for 100 ns) steps. The sampling rate is ~50 Hz. **e,** Cyclability of the transmission change of a photonic switch with 10-nm-thick GST and a 5-μm-long active region at 1,550 nm under multiple Set (3.1 V for 50 μs, 30-μs falling edge) and Reset (7 V for 100 ns) cycles. Each pulse is temporally separated by more than 100 μs to ensure long enough thermal relaxation. The blue and orange shaded areas represent the two-standard-deviation intervals for the amorphous and crystalline states, respectively. aGST (cGST), amorphous (crystalline) GST.

By integrating the photonic switching units on microring resonators (Fig. 4a-b), we can achieve even higher optical contrast from such a compact structure by taking the advantages of both the strong attenuation and optical phase modulation effects of GST. Figure 4c shows the transmission spectra of a microring resonator with a radius of 20 μm integrated with a 3-μm-long switching unit covered with 10-nm-thick GST. After several Set and Reset cycles, the spectra remain the same, indicating reversible phase transitions. As a result of the combined effects of the resonance shift (~0.02 nm/μm GST) and resonance dip (linewidth and depth) change due to the loss modulation (~0.25 dB/μm GST) upon phase transitions (see Supplementary Section 2 for resonance wavelength and loss extraction), a high on-off extinction ratio up to 14.7 dB was achieved near the resonance wavelength. High cyclability with over more than 1,000 phase transitions was also realized (Fig. 4d) in a ring-based switch with the same structure. Note that due to a slight shift between the probe laser wavelength and the ring resonance, the extinction ratio here is not optimal. The relatively large uncertainty of the transmission in the crystalline state can be attributed to the thermal shift of the resonant wavelength and partial crystallization.

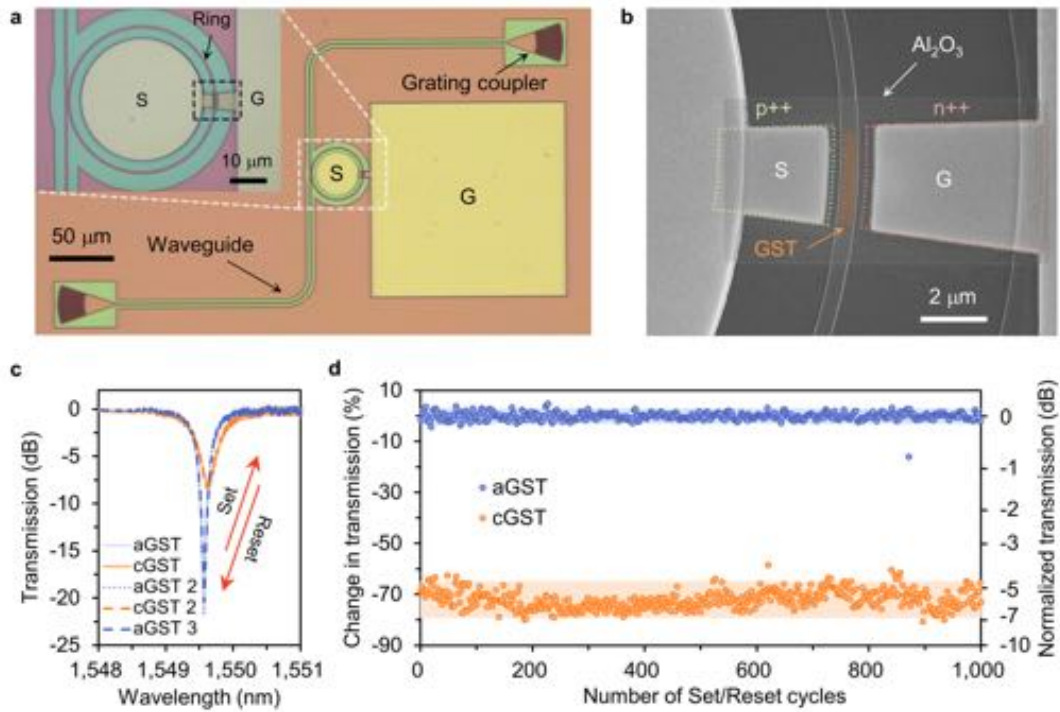

**Figure 4. Photonic switching units on microring resonators. a,** Top-view optical microscope image of the switching unit on an all-pass microring resonator. Inset, enlarged view of the ring. **b,** SEM image of the black dashed area in the inset of **a**. False color is used to highlight the GST (orange). S (G), signal (ground) electrode. p++ (n++), heavily doped p (n)-type silicon region. **c,**

Transmission spectra of a ring switch after a few reversible Set and Reset processes. **d,** Cyclability of the transmission change and normalized transmission of a ring switch at 1,549.57 nm under multiple Set and Reset cycles. Each pulse is temporally separated by more than 0.1 s. The blue and orange shaded areas represent the two-standard-deviation intervals for the amorphous and crystalline states, respectively. Here, the radius of the rings is 20 μm, the gap between the rings and the bus waveguides is 240 nm, the length of the switching units is 3 μm, and the thickness of the GST is 10 nm. Set, a single pulse of 1 V for 100 μs with a falling edge of 60 μs. Reset, a single pulse of 2.5 V for 100 ns. aGST (cGST), amorphous (crystalline) GST.

In summary, we have demonstrated nonvolatile electrically reconfigurable PCM-integrated photonic switches with near-zero additional insertion loss (~0.02 dB/μm) and high endurance (> 1,000 transitions) using in-situ silicon waveguide PIN heaters. By leveraging the remarkable broadband attenuation and optical phase modulation of GST, high extinction ratios are obtained in small footprints (1.25 dB/μm for waveguides and ~15 dB for a microring with a 3-μm-long switching unit). The static nature of the modulation ensures intrinsically high energy-efficiency. We expect that the extinction ratio can be further improved by carefully choosing the power of the applied pulses to achieve complete phase transitions. Multi-level operations will also be possible when the intermediate states are reached via pulse engineering. The insertion loss of the devices can be reduced through thinner electrodes and improved fabrication (Supplementary Section 2). With low-energy, compact, low-loss, and high-cyclability operations at moderate speeds (~10 kHz) as well as the easy-to-deposit property of PCMs and the mature CMOS technology for silicon PIN diodes, our static photonic switching units promise large-scale integration of programmable PICs and pave the way for optical FPGAs.

**Methods**

**Device fabrication.** The photonic switches were fabricated on a commercial SOI wafer with a 220-nm-thick silicon layer on top of a 3-μm-thick buried oxide. The rib waveguides, microring resonators, and grating couplers (with a pitch of 744 nm and a duty cycle of 0.5, inset of Fig. 1b) were defined by EBL (JEOL JBX-6300FS) using positive tone resist (ZEP-520A) and etched by inductively coupled plasma reactive ion etching (ICP-RIE, Oxford PlasmaLab 100 ICP-180) exploiting mixed gas of $SF_6$ and $C_4F_8$. The p++ (n++) regions were defined by another EBL using 600-nm-thick poly(methyl methacrylate) (PMMA) resist and implanted by boron (phosphorus) ions with a dosage of $2 \times 10^{15}$ ions/cm$^2$ and ion energy of 14 keV (40 keV). The ion implantation was conducted at a tilt angle of 7 degrees to misalign with the lattice of silicon and thus achieve uniform deep doping. Subsequently, the chips were annealed at 950 °C for 10 mins to activate the dopants. Before metallization, the surface native oxide was removed by immersing the chips in 10:1 buffered oxide etchant (BOE) for 10 s to ensure Ohmic contact. The metal contacts were then immediately patterned by the fourth EBL using PMMA and formed by electron-beam evaporation (CHA SEC-600) and lift-off of Ti/Pd (5 nm/180 nm) layers. After patterning the windows for the GST by EBL with PMMA, 10-nm or 20-nm GST was deposited onto the chips using a GST target (AJA International) in a magnetron sputtering system (Lesker Lab 18) followed by a lift-off process. The GST and part of the electrodes close to the heating region were then encapsulated by 30-nm-thick $Al_2O_3$ through EBL patterning with PMMA, ALD (Oxford Plasmalab 80PLUS OpAL ALD) at 100 °C, and lift-off. Finally, rapid thermal annealing (RTA) at 200 °C for 10 mins was performed to ensure the complete crystallization of the GST.

**Experimental setup and static measurement.** The photonic switches were characterized by a vertical fiber-coupling setup (Supplementary Section 1). All the measurements were performed under ambient conditions while the temperature of the stage was fixed at ~22 °C by a thermoelectric controller (TEC, TE Technology TC-720) to prohibit the serious thermal shift of the resonators. The input light was provided by a tunable continuous-wave laser (Santec TSL-510) and its polarization was controlled by a manual fiber polarization controller (Thorlabs FPC526) to match the fundamental quasi-TE mode of the waveguides. Focusing grating couplers optimized for ~1,550 nm were employed to couple light into and out of the devices operated at an angle of ~25 degrees. The power of the coupled light was kept to be sufficiently low (< 100 μW) in order to minimize the thermo-optic effects of GST[43] and silicon[44] and avoid causing any phase transition. A low-noise power meter (Keysight 81634B) was used to collect the static optical output from the grating couplers. Based on this setup, we performed the transmission spectrum measurement after each important fabrication step for all the devices including the reference devices without any doping, metal, or GST to assist the insertion loss analysis (Supplementary Section 2) and normalization. To measure the I-V characteristic and conduct electrical switching, electrical signals were applied to the contacts by a pair of DC probes controlled by two probe positioners (Cascade Microtech DPP105-M-AI-S). In particular, the current sweep and voltage measurement were provided by a source meter (Keithley 2450) and the Set and Reset pulses were generated from a pulse function arbitrary generator (Keysight 81160A). The measured I-V curves were used to estimate the power of the applied pulses. By comparing the pulses directly from the pulse generator and those from the probes via a fast oscilloscope (Agilent DSO1022A), no signal distortion was found, suggesting that the probe system has enough response speed for the experiment. Limited by the low control speed of the pulse generator (each pulse takes over 0.1 s), cyclability operations with higher order of magnitude will take hours and thus are not conducted since the setup will get badly misaligned within such long time.

**Transient response measurement.** To obtain the time-resolved response of the photonic switches under the electrical pulses, high-speed low-noise photoreceivers (New Focus 1811 and 1611) were used to measure the dynamic optical signals and the converted output electrical signals were recorded using the oscilloscope triggered by the applied pulses from the pulse generator (see Supplementary Section 1 for the setup). The 125-MHz photoreceiver (New Focus 1811) has enough response speed and appropriate linear operation region to analyze the phase transitions (Fig. 2a-b), pulse amplitude modulation (PAM) effects, and transient response of the rings at different wavelengths (Supplementary Section 3). However, for the analysis of pulse width modulation (PWM) effects, the 1-GHz photoreceiver (New Focus 1611) is necessary. Since its linear operation power is ~1 mW, the output light from grating couplers was first amplified by an optical fiber amplifier (Amonics AEDFA-30-B-FA) and filtered by a narrow-band optical demultiplexer (DEMUX, JDS WD1504D4A-DSC4). The power of the amplified light was then attenuated by a variable optical attenuator (Keysight 81570A) to meet the requirement of linear operation. In this experiment, the wavelength of the input light was fixed at 1549.32 nm in corresponding to the nominal central wavelength of the second channel of the DEMUX.

**Device modeling and simulation.** We developed a coupled electro-thermal two-dimensional (2D) finite-element method (FEM) model using COMSOL Multiphysics to qualitatively simulate the electrical switching of the photonic devices with PIN heaters. In the simulation, a semiconductor module (Semiconductor Interface) based on the Poisson's equation, current continuity equation, and drift-diffusion current density equations[45] was utilized to predict the electric potential, current density, and carrier density distributions in the PIN junctions. A heat transfer module (Heat Transfer in Solids Interface) based on the heat transfer equation $\rho C_p \frac{dT}{dt} = \nabla \cdot (k_{th} \nabla T) + Q_e$ (where $\rho$ is the material density, $C_p$ is the specific heat, $k_{th}$ is the thermal conductivity, and $Q_e$ is the heat source) was exploited to predict the temperature distributions of the whole switching units. The two modules were cross-coupled via the temperature-dependent material parameters (Supplementary Section 4) and the total heat calculated from the semiconductor module including Joule heating and nonradiative recombination heating. The schematic of the model was consistent with the actual device cross-section except that the dopants here were assumed to be uniformly distributed for the sake of simplicity.

In the semiconductor module, the Fermi-Dirac carrier statistics and Jain-Roulston bandgap narrowing model were applied due to the high doping level. The Arora mobility model was used to simulate the effect of phonon/lattice and impurity scattering while the Fletcher mobility model was added to describe the carrier-carrier scattering at high voltage. Trap-assisted recombination and Auger recombination for high bias were also considered in the module. The metal contacts were assumed to be ideal Ohmic and the applied pulses had ideal shapes. All the other external boundaries were electrically insulated.

In the heat transfer module, the infinite element domains were adopted for the left, right, and bottom boundary regions of the model while the convective heat flux boundary condition was used on the surface with a heat transfer coefficient of 5 W/(m²·K). Considering the relative thinness of GST and high operating temperature, thermal boundary resistance (TBR) and surface-to-surface radiation boundaries were utilized. For simplicity without losing generality, the phase transition processes were phenomenologically modeled as that the material properties of GST are weighted sums of those in the amorphous and crystalline states in a small temperature interval (10 K for melting, 100 K for quenching and crystallization) centered at $T_m$ (888 K) or $T_g$ (625 K) of the GST with latent heat (66.81 kJ/kg) and exothermic heat (37.22 kJ/kg) involved[41]. Note that $T_g$ was set to be higher than usual (~423 K) due to the increased $T_g$ at a high heating rate[46].

The mode profiles of the photonic switching units were also modeled based on the same geometry but simulated using a frequency-domain 2D FEM wave optics model through the mode analysis (eigenvalue solver). The carrier density and temperature distributions calculated from the electro-thermal model were employed to determine the complex refractive index of the materials (Supplementary Section 4) during the Set and Reset processes. The transmission of the switching units was then estimated based on the solved effective indices (Supplementary Section 5), qualitatively agreeing with the transient response obtained in the experiment. The discrepancy in specific values is attributed to the simple model of the phase transitions model as well as the uncertainty in the material properties and volume of GST phase transitions. More sophisticated simulations can be conducted by including the kinetic models of melting, vitrification, nucleation, and growth[41] and applying accurate material parameters obtained by material characterizations in the future.

**Acknowledgements**

The research was funded by the SRC grant 2017-IN-2743 (Fund was provided by Intel), Samsung GRO, NSF-EFRI-1640986, and AFOSR grant FA9550-17-C-0017. S.D. acknowledges support from Stanford Non-Volatile Memory Technology Research Initiative (NMTRI). A.M. acknowledges support from Sloan Foundation. Part of this work was conducted at the Washington Nanofabrication Facility / Molecular Analysis Facility, a National Nanotechnology Coordinated Infrastructure (NNCI) site at the University of Washington, which is supported in part by funds from the National Science Foundation (awards NNCI-1542101, 1337840 and 0335765), the National Institutes of Health, the Molecular Engineering & Sciences Institute, the Clean Energy Institute, the Washington Research Foundation, the M. J. Murdock Charitable Trust, Altatech, ClassOne Technology, GCE Market, Google, and SPTS.

**Author contributions**

J.Z., J.K.D., and A.M. conceived the project. J.Z. simulated, designed, and fabricated the devices. J. Z. performed the experiments. Z.F. helped with the reversible switching of PCMs. C. W. and M.L. helped with the transient response measurement. S.Z. and S.D. conducted the annealing simulation and thermal conductivity modeling. P.X. helped with the transmission measurement. S.D. and E.P. helped with GST deposition and process characterization to increase the endurance. A.M. supervised the overall progress of the project. J.Z. wrote the manuscript with input from all the authors.

**Competing financial interests**

The authors declare no potential conflicts of interest.

# Supplementary Information:
# Nonvolatile electrically reconfigurable integrated photonic switch

Jiajiu Zheng[1*], Zhuoran Fang[1], Changming Wu[1], Shifeng Zhu[2], Peipeng Xu[3], Jonathan K. Doylend[4], Sanchit Deshmukh[5], Eric Pop[5], Scott Dunham[1,2], Mo Li[1,2], and Arka Majumdar[1,2*]

[1]Department of Electrical and Computer Engineering, University of Washington, Seattle, WA 98195, USA

[2]Department of Physics, University of Washington, Seattle, WA 98195, USA

[3]Laboratory of Infrared Materials and Devices, Advanced Technology Research Institute, Ningbo University, Ningbo 315211, China

[4]Silicon Photonic Products Division, Intel Corporation, Santa Clara, CA 95054, USA

[5]Department of Electrical Engineering, Stanford University, Stanford, CA 94305, USA

*e-mail: jjzno1@uw.edu; arka@uw.edu

## S1. Experimental setup

The experimental setups as described in the Methods are illustrated in Fig S1. Specifically, the setup in Fig. S1a was used for static measurements including the transmission, I-V characteristic, and cyclability operations. The setup in Fig. S1b was used to measure the transient response of the photonic switching units due to electrical heating including phase transitions, pulse amplitude modulation (PAM), pulse width modulation (PWM), and real-time transmission of microring resonators at different wavelengths (Supplementary Section 3).

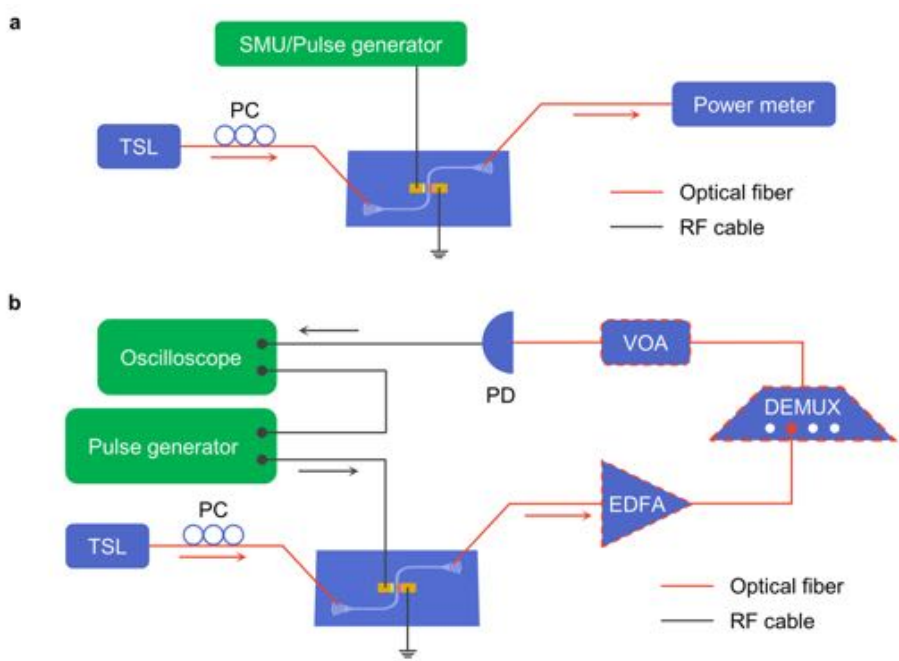

**Figure S1. Experimental setups. a,** Setup for static measurement. **b,** Setup for transient response measurement. The tools framed with dashed red lines were used for measuring the pulse width modulation (PWM) effects (Supplementary Section 3). TSL, tunable laser. PC, polarization controller. SMU, source meter unit. EDFA, erbium-doped fiber amplifier. DEMUX, optical demultiplexer. VOA, variable optical attenuator. PD, photodiode.

## S2. Insertion loss analysis

To analyze the insertion loss (IL) of the photonic switching units, we first performed the mode analysis simulation for the nominal waveguide structures with geometric parameters as outlined in the fabrication processes (see Methods). The refractive indices of the materials used in the simulation can be found in Supplementary Section 4. As listed in Table S1, near-zero additional loss ($6.77 \times 10^{-3}$ dB/μm) is introduced due to the doping and annealing. Metal contacts bring about an IL of ~0.017 dB/μm. Considering that we used relatively thick electrodes (~180 nm) for a good connection near the etching steps and the electrodes were put only 500 nm away from the waveguides to ensure low resistance, a lower IL can be achieved by reducing the thickness of the metal contacts and increasing their distance from the waveguides with improved fabrication. Amorphous GST also introduces a moderate IL of ~0.012 dB/μm. The total IL, defined as the loss in the amorphous state, is thus the sum of the IL from the above sources and is ~0.035 dB/μm. By comparing the loss between the amorphous and crystalline states, we can acquire a theoretical on-off extinction ratio (ER) of ~1.77 dB/μm.

**Table S1. Simulated complex effective index ($n_{eff} - \kappa_{eff}$ i) and attenuation coefficient ($\alpha = 4\pi\kappa_{eff}/\lambda$) of the photonic switching unit after each important fabrication step at 1,550 nm.**

|  | $n_{eff}$ | $\kappa_{eff}$ | $\alpha$ (dB/μm) |
| --- | --- | --- | --- |
| **As-fabricated rib waveguide** | 2.55 | 0 | 0 |
| **After doping and annealing** | 2.55 | $1.92 \times 10^{-4}$ | $6.77 \times 10^{-3}$ |
| **After metallization** | 2.55 | $6.61 \times 10^{-4}$ | $2.33 \times 10^{-2}$ |
| **After sputtering (10 nm aGST)** | 2.58 | $9.98 \times 10^{-4}$ | $3.51 \times 10^{-2}$ |
| **After RTA (10 nm cGST)** | 2.68 | 0.05 | 1.81 |

The IL can also be evaluated through the measured transmission spectra of the waveguides after each important fabrication step. For waveguides, the IL due to a certain fabrication step is calculated as the difference of the transmission differences between the target and reference device (with the same structure but without any doping, metal, or GST) before and after the fabrication step. The reference devices were used to exclude the IL from the grating couplers and alignment variances. As shown in Fig. S2a, the IL induced by the doping and annealing is less than 0.4 dB for the 5-µm-long switching unit, but it is difficult to determine the exact IL because of the ripples from the back reflection and the normalization error from the reference device. A larger IL compared to the simulation is also possible due to extra scattering loss. The IL from metal contacts is ~1 dB which is significantly higher than the theoretical value. This can be attributed to the positional deviation of the electrodes owing to the limited alignment precision. The IL introduced by amorphous GST is ~0.4 dB which is still higher than expected and can be explained by the unwanted scattering loss due to the surface roughness of GST[1]. The total IL is thus around 1.6 dB while the ER is ~1.2 dB. The lower ER compared to the simulation is attributed to material degradation during the fabrication processes.

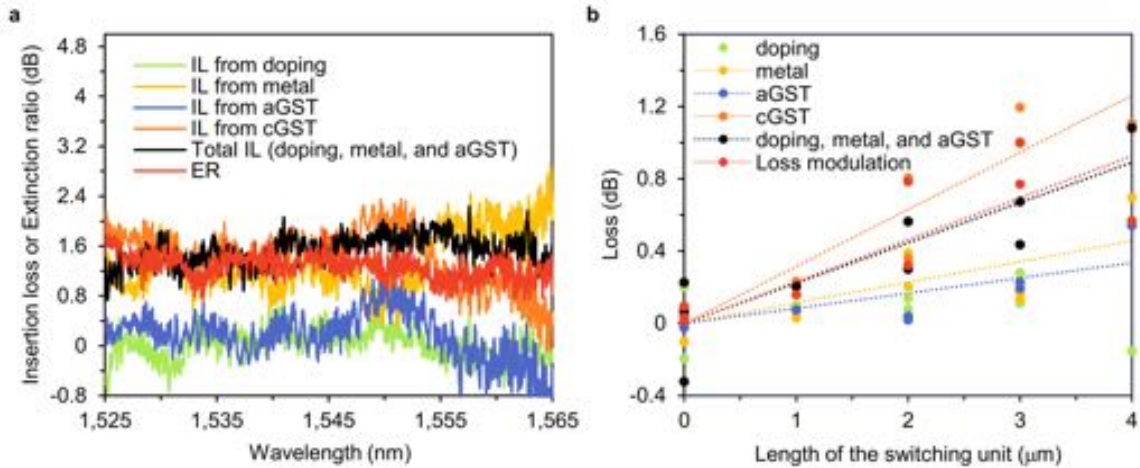

**Figure S2. Insertion loss analysis based on experimental results. a,** Insertion loss (IL) spectra introduced by each important fabrication step and extinction ratio (ER) between two states of a photonic switching unit on a waveguide with 10-nm-thick GST and a 5-µm-long active region (same device as in Fig. 3e). **b,** Loss from each important fabrication step and loss modulation of microring resonators with different lengths of switching units near 1,550 nm. Here, the radius of the rings is 20 µm, the gaps between the rings and the bus waveguides are 240 nm, 270 nm, or 330 nm, and the thickness of the GST is 10 nm. Every dot represents the experiment data of a single device. The dotted lines are the linear fitting of the experiments. aGST (cGST), amorphous (crystalline) GST.

Additionally, IL analysis can be performed by the length-dependent loss statistics from microring resonators. In particular, the quality factors and resonance wavelengths of resonance dips for the rings with different lengths of switching units were first extracted by fitting the measured spectra to a Lorentzian function. The loss introduced by each fabrication step can then be estimated as the difference in the loss of the rings before and after the step through Eq. 1 in Ref. 1. As plotted in Fig. S2b, the loss from each step except doping is linearly proportional to the length of the switching units while the loss due to doping randomly locates near 0 dB, suggesting a negligible additional loss from the PIN junction. As a result, we estimated the IL from metal contacts to be ~0.1 dB/μm, the IL from amorphous GST to be ~0.08 dB/μm, the total IL to be ~0.2 dB/μm, and the IL from crystalline GST to be ~0.3 dB/μm. Therefore, the IL from doping is calculated to be ~0.02 dB/μm and the loss modulation between two states to be ~0.23 dB/μm. These results are comparable with those estimated by the single waveguide.

## S3. Heating dynamics

We studied the heating dynamics of our photonic switching systems by PAM, PWM and microring resonators. During the whole experiment, the GST was kept in the crystalline state pumped by very low pulse energy. For PAM, electrical pulses with a fixed width and variable amplitudes were employed to a switching unit on a waveguide. As presented in Fig. S3a, with the increase of the pulse amplitude, the dynamic change in the transmission increases accordingly. Note that this transmission change originates from the ultrafast free-carrier absorption effect of silicon and the slow thermo-optic effect of GST. The rise time and the dead time (cooling time constant) of the system, however, remain the same, suggesting a pulse amplitude-independent heating and cooling rate. Whereas the heating dynamics does not rely on the pulse amplitude, PWM, where the pulse width was varied but the pulse amplitude was fixed, displays considerably higher cooling rates for shorter pulses (Fig. S3b). This could be intuitively understood that for longer pulses, more energy will get lost into the waveguide and substrate due to the thermal diffusion[2,3], so that a larger heat capacity and a longer thermal time constant are expected leading to a longer dead time. Hence, in order to ensure low energy consumption and high-speed operations, relatively short pulses are preferred. Considering that the temperature gradient in the GST increases significantly with the increase of pulse amplitude[3], pulses with low voltage are also desirable to avoid re-amorphization, ablation, and melting of silicon.

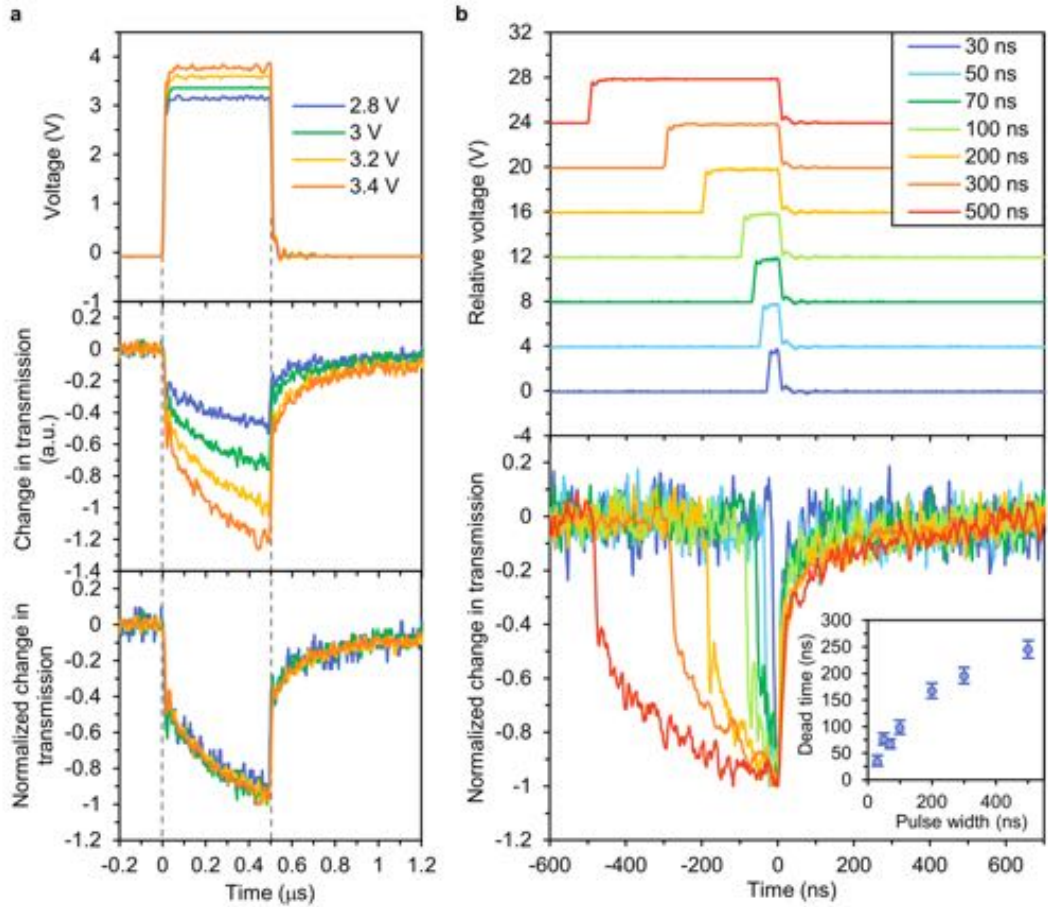

**Figure S3. Heating dynamics of the photonic switching unit on the waveguide. a,** Electrical pulses (upper panel), corresponding transmission response (middle panel) and its normalization (lower panel) of the switching unit at 1,550 nm for pulse amplitude modulation (PAM). **b,** Electrical pulses (upper panel) and normalized transmission response (lower panel) of the switching unit at 1,550 nm for pulse width modulation (PWM). The curves of the electrical pulses with different pulse widths have been vertically offset for clarity. Inset of lower panel, extracted dead time (1/e cooling time) as a function of pulse width with 95% confidence bounds. Here, the length of the switching unit is 5 μm and the thickness of the GST is 10 nm (same device as in Fig. 2). The pulse energy was chosen to be sufficiently low to prevent any phase transition.

From the real-time optical transmission of a switching unit on a microring resonator (Fig. S4), dynamic optical phase modulation can be observed. Here, the probe wavelengths (the inset of Fig. S4) were selected to be gradually shifted from the left side of the resonance to the right side. Due to the ultrafast free-carrier dispersion and absorption effect of silicon, the resonance shift and resonance dip (linewidth and depth) change happen immediately at the beginning and end of the pulse, resulting in a sharp change in the transmission. The slowly varied optical transmission at other times is, however, dominated by the resonance shift due to the thermo-optic effect of silicon. Oscillatory behavior in the transmitted light (Fig. S4b) was also observed at the rising edge of the pulse, which is the signature of the second-order filter and occurs due to the high quality factor ($Q$) of the optical cavity[4]. This behavior is, however, absent at the falling edge owing to the fact that the elevated loss from the GST at higher temperatures renders a low-$Q$ resonator.

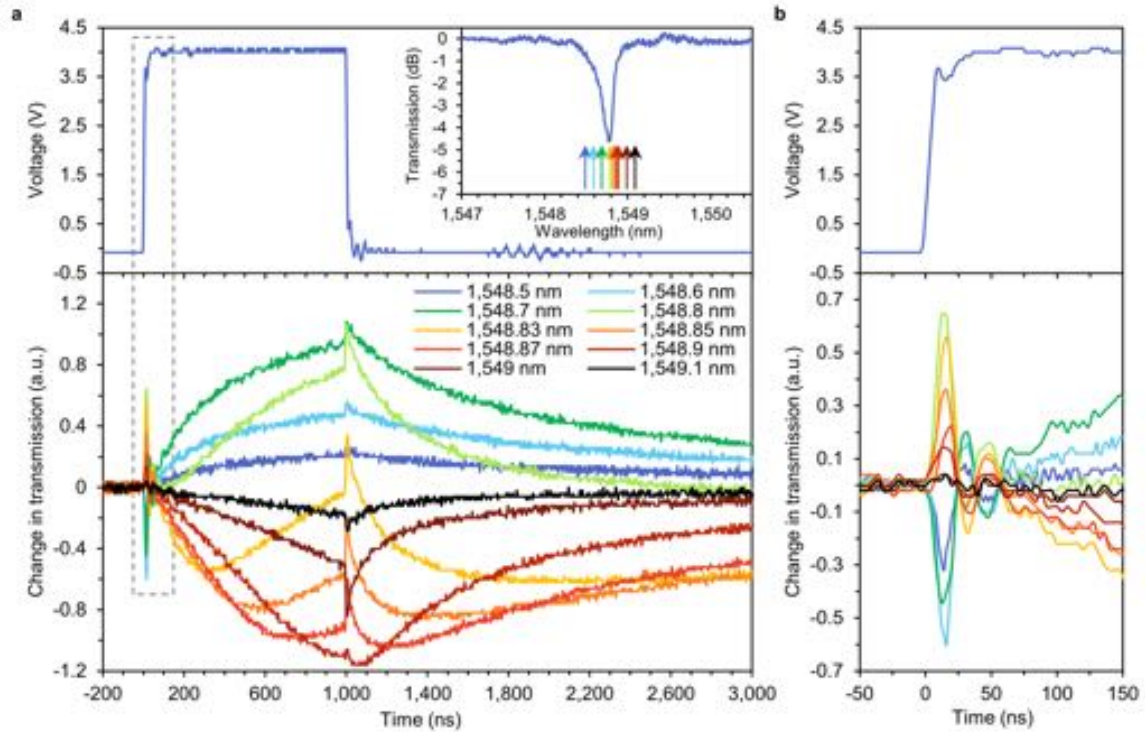

**Figure S4. Heating dynamics of the switching unit on the microring resonator. a,** Electrical pulse (upper panel) and corresponding transmission response (lower panel) of the ring measured at different wavelengths. Inset of upper panel, transmission spectrum of the ring without any voltage applied. The probe wavelengths are annotated in the spectrum by the arrows using the same colors as in the lower panel. **b,** Zoom-in inspection of the grey dashed area in **a**, indicating a combination of the free-carrier effect and cavity oscillatory behavior in response to the step signal. Here, the radius of the ring is 20 μm, the gap between the rings and the bus waveguides is 330 nm, the length of the switching unit is 2 μm, and the thickness of the GST is 10 nm. The pulse energy was chosen to be sufficiently low to prevent any phase transition.

## S4. Material parameters used in simulations

Table S2 lists the main material parameters used in our simulations. Note that, thermal boundary resistance (TBR) was applied to all the internal boundaries and depends both on the temperature and phase of GST[5]. As the GST is partially crystallized during the phase transitions, the complex refractive index of GST was approximated from the effective permittivity equation based on an effective-medium theory[6] where the complex refractive index of the amorphous and crystalline GST and the degree of crystallization from the electro-thermal model were utilized.

**Table S2. Main material parameters used in simulations.**

|  | $n$ | $\kappa$ | $\beta$ (K$^{-1}$) | $\gamma$ (K$^{-1}$) | $k$ (W m$^{-1}$ K$^{-1}$) | $C_p$ (J kg$^{-1}$ K$^{-1}$) | $\rho$ (kg m$^{-3}$) |
|---|---|---|---|---|---|---|---|
| Si | 3.4777 + $\Delta n$[7] | $\Delta \kappa$[7] | $1.8 \times 10^{-4}$ | — | $k(T)$[8] | $C_p(T)$[9] | 2,329 |
| SiO$_2$ | 1.444 | 0 | — | — | $k(T)$ from COMSOL | $C_p(T)$ from COMSOL | 2,200 |
| Al$_2$O$_3$ | 1.5886 | 0 | — | — | 1.9[10] | $C_p(T)$[11] | 3,100 |
| Pd | 3.1640 | 8.2121 | — | — | 71.8 | 244 | 12,023 |
| aGST | 3.8884[1] | 0.024694[1] | $1.1 \times 10^{-3}$[12] | $4.1 \times 10^{-4}$[12] | 0.19[13] | 213[14] | 5,870[15] |
| cGST | 6.6308[1] | 1.0888[1] | $-2.2 \times 10^{-4}$[12] | $1.56 \times 10^{-3}$[12] | $k(T)$[13] | 199[14] | 6,270[15] |

$n$, refractive index. $\kappa$, extinction coefficient. $\beta$, thermo-optic coefficient for refractive index. $\gamma$, thermo-optic coefficient for extinction coefficient. All the optical parameters are for 1,550 nm. $k$, thermal conductivity. $C_p$, heat capacity at constant pressure. $\rho$, density. $T$, temperature. $\Delta n$ and $\Delta \kappa$ are dependent on carrier densities of electrons and holes and can be calculated from Eq. 6.3 and Eq. 6.4 in Ref. 7.

## S5. Calculation of real-time optical transmission

As described in the Methods, we calculated the real-time optical transmission of the switching unit (Fig. S5) according to the effective index change based on the simulated carrier density and temperature distributions from the electro-thermal model during the Set and Reset processes. The applied electrical pulses have the same power and width as used in Fig. 2. From the calculation, a similar steep change in the transmission at the sharp edges of the pulses can be observed, confirming the ultrafast free-carrier absorption in silicon. We can also find gradually varied transmission regions during the heating and cooling processes reflecting the thermo-optic effect of GST. However, the abrupt change of the transmission at around 10 μs for Set and 200 ns for Reset, where the crystallization (nucleation and growth) and vitrification happen, respectively, was not likewise found in the experiment. This discrepancy can be mainly attributed to the simple model of the phase transitions and the uncertainty in the material properties. The actual fraction of the GST undergoing phase transitions in the experiment is also not determined. Therefore, the calculated large change in the transmission might be over-estimated.

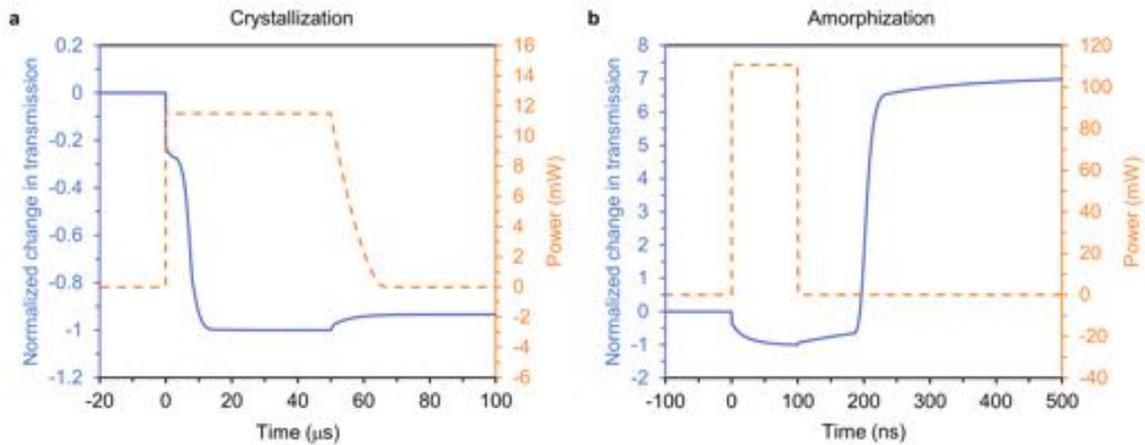

**Figure S5. Real-time optical transmission during Set and Reset processes. a,b,** real-time power of the applied electrical pulse (dashed line) and corresponding normalized change in optical transmission of the simulated switching unit at 1,550 nm (solid line) for crystallization (Set, **a**) and amorphization (Reset, **b**). Here, the length of the switching unit is 5 μm and the thickness of the GST is 10 nm.